# Planning a Reference Constellation for Radiometric Cross-Calibration of Commercial Earth Observing Sensors

**Sreeja Nag[1], Philip Dabney[2], Vinay Ravindra[1], and Cody Anderson[3]**

[1]NASA Ames Research Center, Bay Area Environmental Research Institute, Moffet Field, CA, USA
[2]NASA Goddard Space Flight Center, Greenbelt, MD, USA
[3]United States Geological Survey, Sioux Falls, SD, USA
Sreeja.Nag@nasa.gov

**Abstract**

The Earth Observation planning community has access to tools that can propagate orbits and compute coverage of Earth-observing imagers with customizable shapes and orientation, model the expected Earth Reflectance at various bands, epochs and directions, generate simplified instrument performance metrics for imagers and radars, and schedule single and multiple spacecraft payload operations, including agile re-orientation of satellites. We are working toward integrating existing tools to design a planner that allows commercial small spacecraft to assess the opportunities for cross-calibration of their sensors against current stable, flagship reference sensors. The planner would take a suite of inputs: specifications of the satellite (and instrument) to be calibrated, specifications of the reference instruments, sensor stability that defines the planning horizon, allowable latency between calibration measurements, allowable differences in viewing and solar geometry between calibration measurements, etc. The planner would output cross-calibration opportunities for every reference-target pair as a function of flexible user-defined parameters, which could then be used by an automated ground scheduler, or even a real-time onboard scheduler. We use a preliminary version of this planner to inform the design of a constellation of transfer radiometers that can serve as stable, radiometric references for commercial sensors to cross-calibrate with. We propose such a constellation for either vicarious cross-calibration using pre-selected sites, or top of the atmosphere (TOA) cross-calibration globally. Results from the calibration planner applied to a subset of informed architecture designs show that a 4-sat constellation provides multiple calibration opportunities within half a day's planning horizon, for Cubesat sensors deployed into a typical ride-share orbits. While such opportunities are available for cross-cal image pairs within 5 deg of solar or view directions, and within an hour (for TOA) and less than a day (vicariously), the planner allows us to identify many more by relaxing user-defined restrictions, based on their understanding of their instrument stability or cross-cal algorithmic performance.

## Introduction

Earth observation activities have undergone a significant change in concept of operations in the last decade with the emergence of small satellites (e.g., CubeSats) provisioned with high-quality sensors. Once sensor technologies are validated via NASA's incubator programs for components [e.g. Computational Reconfigurable Imaging Spectrometer (Sullenberger et al. 2017)], full instrument systems [e.g. Compact Midwave Imaging Sensor (Kelly et al. 2010), Raincube (Peral et al. 2015)] or constellation [e.g. TEMPEST-D (Reising et al. 2016)] demonstrations, they have been commercialized via large-scale deployment for continued, global products [e.g. visible imagery by Planet Labs (Doan et al. 2017), radar imagery by Capella Space]. Constellations of small satellites in large numbers, especially with platform agility, enable data acquisition at temporal frequencies and spatial scales that were previously not possible (Nag, Li, and Merrick 2018; Shao et al. 2018), thereby creating actionable information for scientific discovery and exploration guidance. However, the radiometric stability of such small platforms is not comparable to NASA's flagship missions, because they typically use lunar, deep space or vicarious calibration, instead of carrying heavy blackbody sources or noise diodes.

Innovative inter-spacecraft and inter-instrument calibration are required to fully exploit data gathered from multiple locations at different or identical times. Vicarious calibration of sensors using pre-identified, relatively invariant ground targets is a promising method. However, it is restricted by the (A) lack of calibration automation for hundreds of sensors, (B) unavailability of frequent cross-calibration ground site opportunities, and (C) latency due to ground-based processing of images to determine calibration parameters. Process automation has begun to now address [A]: Planet has been able to calibrate its Flock 3p at one satellite per day, and its, and Flock 2k at two satellites every three days, with 75% and 90% of calibration activities proceeding without operator intervention, respectively (Leung et al. 2018). The processes of calibration, correction, stability monitoring, and quality assurance needs to be underpinned and evidenced by comparison with (ideally) highly calibrated in-orbit reference instruments (Gyanesh

Chander et al. 2013). This comparison of physical meaning of information collected by the reference instruments and the relatively unstable, smaller sensors through robust Système International d'unités traceable calibration and validation (Cal/Val), needs them to collect images at the same place in as similar viewing conditions as possible, therefore problem [B]. Problem [C] can be mitigated in part by increasing the number of automated ground stations to support cross-calibration, and onboard processing of calibration parameters along with inter-satellite communication of the parameters and/or image features. Scheduling such activities can leverage the concept of operations for reactive imaging, where in the targets would be calibration targets instead of coordinates of transient phenomena – example framework in (Nag et al. 2019).

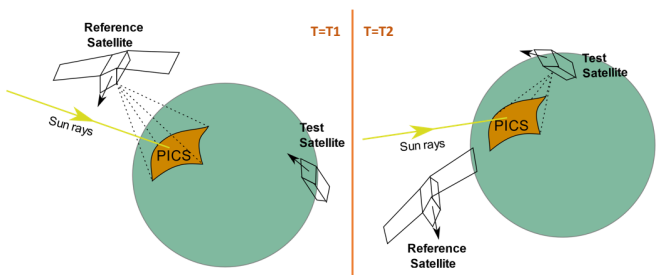

Figure 1—Cross calibration using two image collections by the reference and test satellites, at T=T1 and T=T2 respectively, of pre-identified vicarious calibration sites, e.g. from the database of Worldwide Pseudo Invariant Calibration Sites or PICS (G. Chander et al. 2007)

This paper proposes the use of a constellation of stable, agile transfer radiometers as a reference to enable cross-calibration of smaller sensors, summarizes a planning tool to identify the baseline design of such a constellation, and demonstrates in simulation the increased number of calibration opportunities if such a constellation were deployed. If the sensors are agile i.e., capable of off-nadir pointing in all 3 degrees of freedom (3-DOF), they can attempt to match the viewing geometry of the reference satellite's images within the constraints of orbital mechanics. By employing a novel calibration technique using the top of the atmosphere (TOA), instead of land or oceanic sites, it may be possible to increase calibration opportunities, due to more crossover points of the swaths, albeit at the cost of shorter time windows, owing to the atmosphere being more unstable than land.

Our developed planning tool can help design a reference constellation that will maximize the calibration opportunities for a generic commercial test satellite. Figure 1 shows a cartoon of a hypothetical collection of images at T1 and T2, of one of the Worldwide Pseudo Invariant Calibration Sites/PICS, that will later be used to calibrate the test satellite (e.g. Planet Labs' Dove) against the in-orbit reference (e.g. Landsat 8, 9, VIIRS). As summarized in Figure 2, the tool allows evaluation of various architectures of reference constellations in terms of the calibration opportunities they provide, as a function of allowable range of time difference ($\Delta Tsite$), solar zenith angle ($\Delta SZA$) and view zenith angle ($\Delta VZA$) between the image collections, and required time horizon for calibration ($\Delta Tstab$). $\Delta Tsite$ is dependent on the stability of the selected vicarious site (how long can we assume a PICS site to be invariant?), while $\Delta Tstab$ is dependent on the absolute stability of the sensor (how often does it need cross-cal?). The tool uses an existing orbit propagator and coverage calculator (Orbit Module in Figure 2), namely NASA GSFC's General Mission Analysis

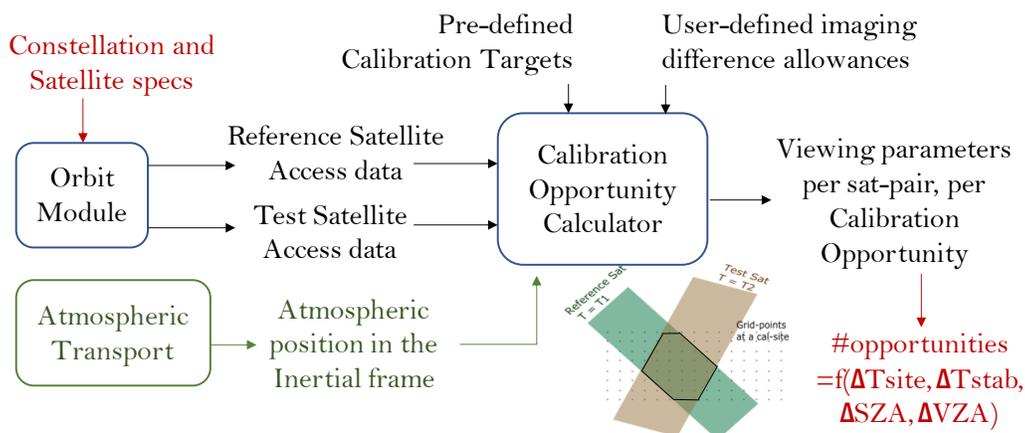

Figure 2 – Process overview of computing cross-calibration opportunities within the calibration planner, leveraging modules developed within the Tradespace Analysis Tool for Constellations (TAT-C). An additional Atmospheric Transport module is needed for computing TOA calibration opportunities.

Tool or GMAT to compute the access intervals for the reference or test satellites, given the coordinates of any number of global calibration points. It uses a modified version of the Instrument Module from the Tradespace Analysis Tool for Constellations or TAT-C (Nag, Hughes, and Moigne 2016; LeMoigne et al. 2017) to compute the solar and view geometry parameters of every access event.

Calibration opportunities between any pair of satellites are computed as the triple overlap of two satellite swaths (brown and green overlay in Figure 2's inset) with the ground calibration site location (grid points in Figure 2's inset) – within user-defined allowed differences in time of collection, solar and view geometry, and calibration planning horizon. The tool also allows us to set the maximum absolute VZA or SZA, to limit the Signal to Noise Ratio (SNR) or the calibrated images.

## Current State of Cross-Calibration

The current state of art for commercial Cubesats is to cross-calibrate against flagship missions over pre-identified relatively invariant sites (Wilson et al. 2017), either by matching basic parameters like brightness temperatures or radiance-dependent products like NDVI (Houborg and McCabe 2018). We use 48 PICS locations (G. Chander et al. 2007) as an example in this article, but the planning tool can accommodate any other ground calibration. Each calibration region is characterized by a 250km x 250 km rectangle, with 200 evenly spaced gridpoints. We use the following 6 flagship instruments as current reference satellites in this article, however can accommodate others like VIIRS or MODIS easily.

1. Operational Land Imager (OLI) on Landsat-8; 15deg cross-track FOV, 710 km SSO
2. OLI on Landsat-7 (which will be replaced by Landsat-9 next year); 15deg cross-track FOV, 710 km SSO
3. Multi-Spectral Imager (MSI) on Sentinel-2A; 20.6deg cross-track FOV, 788 km altitude SSO
4. MSI on Sentinel-2B; 20.6deg cross-track FOV, 788 km altitude SSO
5. Ocean and Land Colour Instrument (OLCI) on Sentinel-3A; 68.6 deg FOV tilted at 12.6deg away from the Sun, 802 km altitude SSO
6. OLCI on Sentinel-3B; same specs as Sentinel-3A

For the test satellites, we consider three Planet Labs' Dove satellites, one as deployed from the International Space Station (ISS) and two in sun synchronous orbits (ISS) in flock 2P and 3P, trailing Landsat 8 and 7 respectively. Since most commercial Cubesats are deployed from rideshare opportunities, the ISS and morning-afternoon SSO orbits represent the most common launches and orbits. The FOV of the sensors is 2 deg x 3 deg. Figure 3 shows the number of calibration opportunities currently available to a commercial Cubesat with off-nadir pointing of 27.5deg cross track, within a 48 hour mission simulation period (max $\Delta T_{stab}$). As the maximum allowed time difference increases there are more cross-cal opportunities. Since all the reference sats are in SSOs, the SSO test sats (hollow and filled circles) have two easily distinguished buckets of cross-cal opportunities around the ~2 hour $\Delta T_{site}$ and ~24 hour $\Delta T_{site}$. The plateaus indicate no new opportunities, in spite of relaxing the $\Delta T_{site}$ requirements.

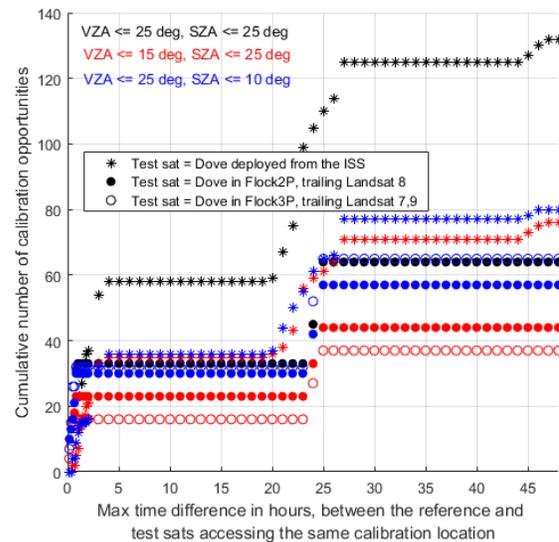

Figure 3—Total number of calibration opportunities available for 3 test satellites, with only cross-track maneuverability, when calibrating against 6 flagship NASA and ESA missions (Landsat 7 and 8, Sentinel 2A, 2B, 3A and 3B) as a function of maximum allowed time gap between the imaging opportunities, and maximum solar and view zenith angles. Absolute stability of the test sat sensor is assumed to last 48 hours.

In the current state of art, the *automated calibration scheduling algorithm* for the test satellite should be to periodically re-evaluate its radiometric stability and assess the horizon of cross-calibration. It should then use the output data from the planning tool to determine the cross-cal opportunities within that horizon, and select them so as to minimize the cross-cal estimation errors of in-house algorithms with $\Delta VZA$ and $\Delta SZA$ as variables. For example, if the PICS invariance is 5 hours and the stability threshold of a test Cubesat in Flock2P is violated in 48 hours, it has 22 opportunities to select from (filled red circles) – the optimum choice is expected to minimize the cross-cal errors for a pair of images that are at most $\Delta VZA=15$ deg and

ΔSZA=25 deg apart. If the Cubesat sensor quality is such that the stability threshold is violated within 6 hours and the ΔSZA difference reduced to 10 deg, there is only 1 cross-cal opportunity (see Figure 4). Accounting for the variability caused by epoch time and exact orbits of the test sat, it is very likely that such angular restrictions will result in no opportunities at all.

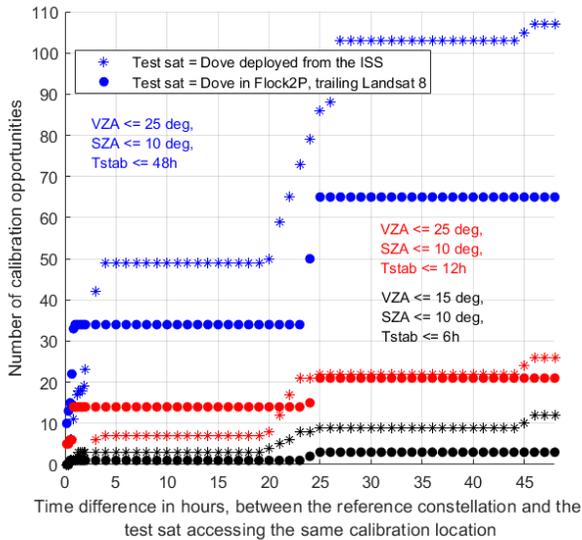

Figure 4—Total number of calibration opportunities available for 2 test satellites, with 3-DOF maneuverability, when calibrating against 6 flagship missions as a function of maximum allowed time gap between the imaging opportunities, and maximum solar and view zenith angles. Absolute stability of the test sat sensor is shown to vary from 6 to 48 hours.

If each satellite's attitude is maneuverable in 3-DOF, i.e. capable of off-nadir pointing within a conical field of regard (FOR) of 55 deg, it may further try to match viewing geometry and increase calibration opportunities. Figure 4's blue curves are the same simulated plan as Figure 3's blue curves, except with 3-DOF maneuverability, therefore 10-20% opportunities. Figure 4 also shows the dependence on planning horizon as determined by test sensor $\Delta T_{stab}$ – more unstable sensors need more frequent calibrations, and have fewer options. The black curves show that, in spite of the extra agility, restricting angular differences still leaves just one opportunity. Moreover, the synchronous orbits cause the opportunities to be clustered in a small spread of ΔTsite and ΔSZA. To allow the lower quality commercial sensors reasonable cross-cal options, special reference satellites will need to complement the current flagship ones.

## Constellation of Reference Calibrators

Deploying a government-provided constellation of transfer radiometers (TR) will increase the number of reference satellites in orbit, and allow more opportunities for cross-calibration. The TR concept for imagers is much like the CLARREO Pathfinder mission, that enables highly accurate decadal change observations by a solar spectrometer, traceable on-orbit to SI standards (Feldman et al. 2011). The TRs would be stable to 0.5% over the course of 2 weeks, as cross-calibrated against the flagship missions (e.g. Landsat 7 and 8, Sentinel 2A, 2B, 3A and 3B instruments in the previous section).

This section will compare the performance of the various TR constellation architectures, as evaluated by the planning tool in Figure 2. While the tool serves to plan in-orbit operations, the input specs may be searched to optimize the output opportunities. While a full constellation designer such as TAT-C may be used for this purpose, a subspace of informed options is analyzed here. For example, the clustering of calibration opportunities for SSO references, especially in SZA, indicates that the TR constellation should not be in SSO. To maximize coverage of invariant ground spots, which are primarily in tropical regions (and Antarctica, but cannot calibrate visible imagers in winter), a ~45 deg inclination was chosen. To keep the mass and size within the ESPA-class and cost low, the FOV of the TRs were restricted to 2 deg x 3 deg. The spatial resolution is expected to be high, and does not put any bounds on the altitude. Instead, a low altitude of 450 km was chosen to maximize the orbital velocity, therefore revisit rates to calibration sites and crossover points with any orbital planes. The TR satellites are expected to be agile, and will be able to point off-nadir in cross and along track in order to recreate the full extent of any Landsat, VIIRS or MODIS images, so that cross-calibration with the flagship missions is possible.

Six constellation architectures with up to three orbital planes, and one to three satellites per plane were assessed (Table 1), in terms of cross-calibration opportunities against an ISS-deployed and an SSO-deployed test satellite. Results are quite different between ground cross-cal and TOA cross-cal, for either test satellite.

Table 1–Topology of constellation architectures. All satellites are in circular orbits, at 450 km altitude, 45 deg inclination. The initial RAAN of the planes are arranged at maximum separation from the Landsat and Sentinel planes, however the separations will periodically oscillate over mission lifetime, due to unequal J2 precession rates.

| Arch# | # sats | # planes |
|---|---|---|
| 1 | 1 | 1 |
| 2 | 2 | 1 |
| 3 | 3 | 1 |
| 4 | 4 | 2 |
| 5 | 6 | 2 |
| 6 | 6 | 3 |

**Vicarious Cross-Calibration using Selected Sites**

The TR constellation is expected to be agile and re-orient to image the vicarious calibration sites. The commercial sat operators are expected to know the access times (and some information on viewing geometries) of these reference images, so that they may plan for the re-orientation of their assets to capture appropriate images for cross-calibration. Figure 5 shows the number of opportunities as a function of constellation architecture and the assumed invariant time of the cal-sites, up to a maximum of 2 days. ΔVZA was restricted to the FOR of 27.5 deg, ΔSZA was unrestricted and ΔTstab stable through 2 days. Figure 6 shows the results from the same planning simulation, except with the ISS deployed Cubesat instead of an SSO one. ISS orbits have far more opportunities than SSO orbits, and their variability with respect to the planning epoch is fairly large. For example, the single plane TR constellations result in more opportunities for ISS than their two plane versions (2 sats does better than 2x2 sats, 3 sats does better than 2x3 sats) because of the way the one vs. two orbital planes are initialized with respect to the PICS; as the planes rotate, performance is expected to caper off.

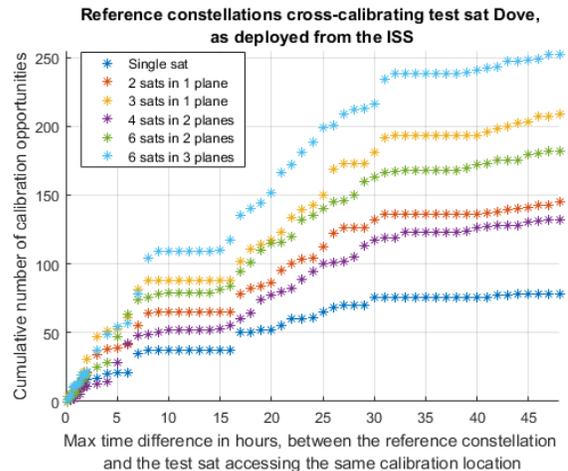

Figure 6– Total number of PICS calibration opportunities available for an ISS-orbit test satellite, when calibrating against different architectures of reference calibrators, as a function of maximum allowed time gap between the imaging opportunities.

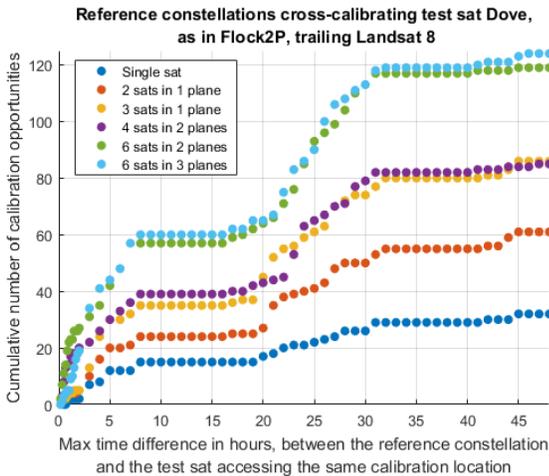

Figure 5– Total number of PICS calibration opportunities available for an SSO-orbit test satellite, when calibrating against different architectures of reference calibrators, as a function of maximum allowed time gap between the imaging opportunities.

The 3 and 4 sat TR constellations provide SSO sats similar number of opportunities, for varying ΔTsite. The performance for varying ΔVZA and ΔSZA for the 4-sat architecture is seen in Figure 7, and this is in addition to what the flagship references can provide. The planning tool therefore allows commercial operators to predict the occurrence of cross-cal options within their sensor stability planning horizon, and selection from these occurrences by minimizing algorithm error due to image acquisition Δ. Each op-

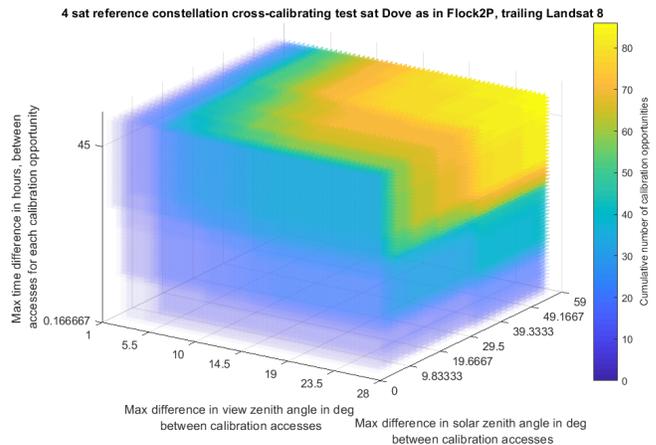

Figure 7– Total number of PICS calibration opportunities available for an SSO-orbit test satellite, when calibrating against a 4-satellite constellation (purple curve of Figure 5) of reference calibrators, as a function of maximum allowed time gap, SZA and VZA difference between the imaging opportunities.

Coverage of the selected constellations, as defined by number of opportunities to visit any PICS grid point within a given revisit time, is seen to be mainly driven by the number of satellites, irrespective of how they are arranged. However, the cross-cal opportunities with specific mean anomalies of specific orbits are sensitive to those specifics, and this is especially true when the cross-cal is tied to specific geographic locations. To explore the relaxation of the site criteria to enable a generically applicable TR constellation, we then propose global calibration using the top of the atmosphere (TOA).

**Top of the Atmosphere Cross-Calibration**

Since the TOA is global, the TR constellation is expected to be agile only to cross-cal against the full images of flagship references like VIIRS using the narrow FOV TR sensor. During normal operations, the TR sensors are expected to be pointed nadir and image a narrow swath. The commercial sat operators are expected to know the access times (and some information on viewing geometries) of the TR images, so that they may plan for the re-orientation of their assets to capture appropriate images for cross-calibration. A cross-cal opportunity then corresponds to only the overlap of the TR FOV ground track with the test sat FOR ground track. The lack of the third overlap layer (e.g. PICS grid points), increases the number of opportunities per orbital period of the test sat, however, since the atmospheric conditions are more dynamic than land, the allowed time difference between cross-cal images (ΔTstab) is much more than an order of magnitude less. Figure 8 shows the increasing number of opportunities with allowable time difference (ΔTstab<=2 hours) for different constellation architectures, as available to an SSO test sat. Figure 9 shows the same for an ISS deployed test sat. As mentioned before, the planner shows that the 6 satellite case does similarly, whether arranged in 2 or 3 planes (blue and green curves). This makes a good case to launch as secondary payloads, as launches become available, since the assets don't have to be in the same plane. It will also be possible to flexibly scale up TR capability with time.

TOA cross-cal opportunities for an SSO test sat is very low for TR constellations in a single plane (yellow and orange curves in Figure 8) – a single shot in an hour. While this behavior is dependent on the time of year, there will be spans when no options are available in an hour. The ISS test sat shows a steady increase in opportunities with TR sat number. Eitherway, the 4 sat TR constellation (purple curves) shows similar performance in both cases, as was also observed in the PICS simulations. The results may change as more altitudes and inclinations for the TRs are explored, however given the unpredictable arrangement of commercial Cubesats in space, the 2x2 constellation is chosen as a planning baseline to compare the capabilities of a TR constellation using TOA or PICS with current options.

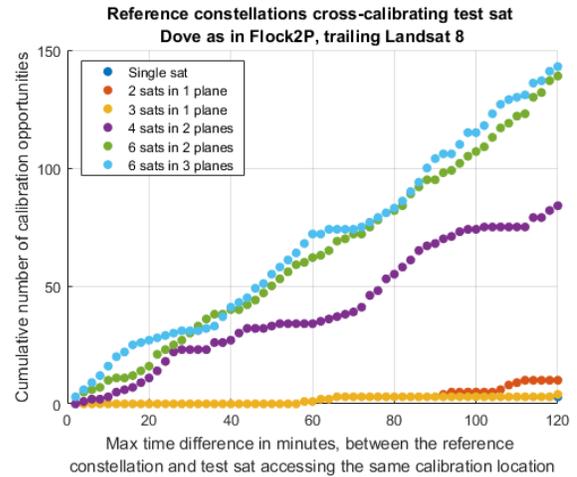

Figure 8– Total number of TOA calibration opportunities available for an SSO-orbit test satellite, when calibrating against different architectures of reference calibrators, as a function of maximum allowed time gap between the imaging opportunities.

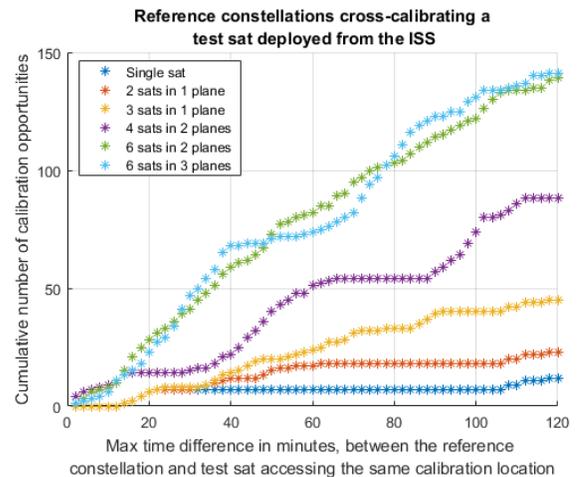

Figure 9– Total number of TOA calibration opportunities available for an ISS-orbit test satellite, when calibrating against different architectures of reference calibrators, as a function of maximum allowed time gap between the imaging opportunities.

ΔVZA is not expected to be as much a plan driver in the TOA case because the test sat can take a nadir image at a pre-known crossover point, similar to the TR's nadir image. The dependence on ΔSZA for the 4-sat constellation is seen in Figure 10 and Figure 11. While the number of opportunities are similar for both test sats, sensitivity to ΔSZA, ΔTsite and ΔTstab has some differences, which the calibration planning tool can help optimize. The contours show the planning period within which opportunities need to be identified, based on how long it takes for a test sensor deviate from a pre-defined threshold stability. Test sensors orbit notwithstanding, a 4-sat TR constellation allows at least 10 cross-cal opportunities within a 12 hour stability horizon for <5 deg of ΔSZA and <1 hour of ΔTsite (TOA) variability between the image pairs. In comparison, the 3 sat constellation provides <10 opportunities for an ISS test sat and 2 opportunities for an SSO test sat, for ~10 deg of ΔSZA and ~1 hour of ΔTsite (TOA) between the image pairs, over a 2 day planning horizon. If sensor stability is around a day, they will need cross-cal algorithms that allow relaxing their view difference requirements.

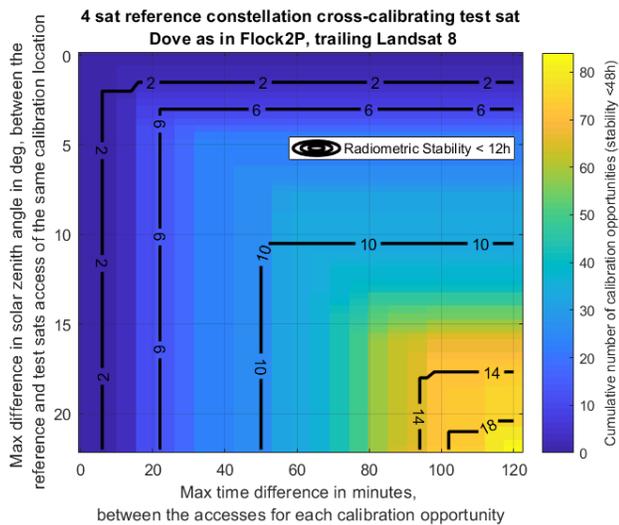

Figure 10– Total number of TOA calibration opportunities available for an SSO-orbit test satellite, when calibrating against a 4 satellite constellation of reference calibrators (purple curve in Figure 8), as a function of maximum allowed time gap and SZA difference between the imaging opportunities.

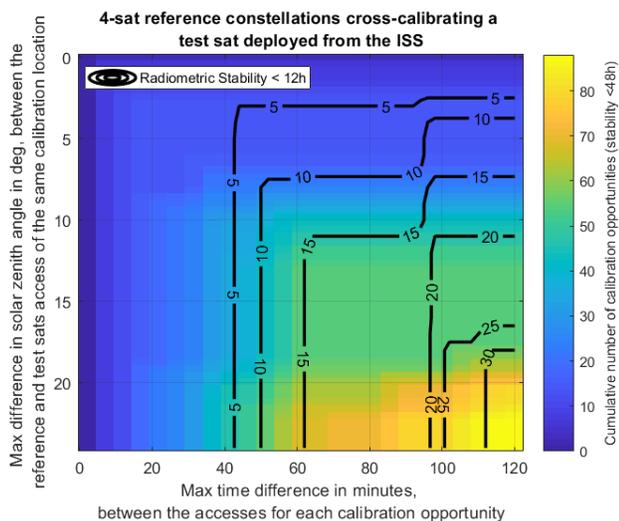

Figure 11– Total number of TOA calibration opportunities available for an SSO-orbit test satellite, when calibrating against a 4 satellite constellation of reference calibrators (purple curve in Figure 8), as a function of maximum allowed time gap and SZA difference between the imaging opportunities.

# Acknowledgements

This work has been supported by the NASA Advanced Information Systems Technology grant awarded by NASA's Earth Science Technology Office, and NASA's Sustainable Land Imaging – Technology program.

## Appendix

The geographic distribution of cross-calibration opportunities, i.e. the crossover sites, for the TOA between the proposed constellation of four reference calibrators and two test satellites, with no restriction to VZA or SZA or time between accesses over the 48-hour simulation is shown below. The TOA method is likely going to be most effective for a maximum time difference of 2 hours, therefore the plots in the article have restricted it to that value.

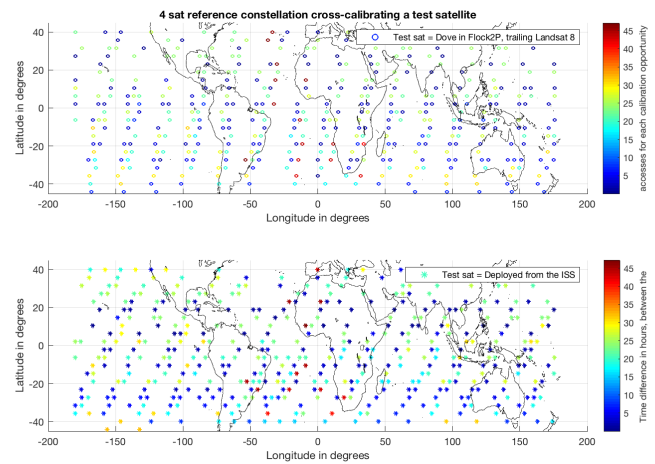